\documentclass[11pt]{article}
\def\beqar {\begin{eqnarray}}
\def\eeqar {\end{eqnarray}}
\def\beq {\begin{equation}}
\def\eeq {\end{equation}}
\def\ra {{\rangle}}
\def\la {{\langle}}
\def\half {{\textstyle{1\over 2}}}
\def\Tr {{\rm Tr}}
\def\tr {{\rm tr}}

\def\del {{\partial}}

\def\d {{\delta}}
\def\l{{\lambda}}

\def\bz {\bar{z}}

\def \A {{\cal A}}
\def \D {{\cal D}}

\def\no2 {{\textstyle{n\over 2}}}

\begin{document}

\begin{titlepage}
\null\vspace{-62pt}

\pagestyle{empty}
\begin{center}
\rightline{}
\rightline{}

\vspace{1.0truein} 
{\Large\bf Bosonization of the lowest Landau level in arbitrary dimensions: }\\
\vskip .2in
{\Large\bf edge and bulk dynamics}\\
\vskip .3in
\vspace{.5in}DIMITRA KARABALI \footnote{E-mail address: dimitra.karabali@lehman.cuny.edu} \\
\vspace{.2in}{\it  Department of Physics and Astronomy\\
Lehman College of the CUNY\\
Bronx, NY 10468}\\
\end{center}
\vspace{0.5in}

\centerline{\bf Abstract}

We discuss the bosonization of nonrelativistic fermions interacting with non-Abelian gauge fields in the lowest Landau level in the framework of higher dimensional quantum Hall effect. The bosonic action is a one-dimensional matrix action, which can also be written as a noncommutative field theory, invariant under $W_N$ transformations. The requirement that the usual gauge transformation should be realized as a $W_N$ transformation provides an analog of a Seiberg-Witten map, which allows us to express the action purely in terms of bosonic fields. The semiclassical limit of this, describing the gauge interactions of a higher dimensional, non-Abelian quantum Hall droplet, produces a bulk Chern-Simons type term whose anomaly is exactly cancelled by a boundary term given in terms of a gauged Wess-Zumino-Witten action. 

\vskip .1in
\vskip .5in

\baselineskip=18pt

\end{titlepage}

\hoffset=0in
\newpage
\pagestyle{plain}
\setcounter{page}{2}
\newpage

\section{Introduction}

Quantum Hall effect in higher dimensions and different geometries has become a topic of recent research interest \cite{HZ}-\cite{poly1}. The higher dimensional generalization exhibits features similar to the two-dimensional case, such as incompressibility and gapless edge excitations, among other things.  In a series of papers \cite{KN1}-\cite{KA} we generalized the original Zhang-Hu construction of QHE on $S^4$ to arbitrary even dimensions by formulating the quantum Hall effect on the complex projective spaces ${\bf CP}^k$. Within this framework we also introduced a bosonization approach for nonrelativistic fermions in higher dimensions. The possible excitations of the LLL fermionic system in the presence of a confining potential are particle-hole excitations, which can, in principle, be described in terms of bosonic degrees of freedom. Using the quantum density matrix formulation we derived an exact bosonic action describing these excitations \cite{KN2, KN3}. The bosonic action is given in terms of one-dimensional $(N \times N)$ matrices acting on the $N$-dimensional LLL single particle Hilbert space, which can be further expressed, using the star-product formulation, as an action of a  noncommutative field theory. 

In \cite{KA} we extended the bosonization method outlined in \cite{KN2, KN3} in the case of electromagnetic,  $U(1)$,  gauge interactions. ${\bf CP}^k$ however is a space which also admits a non-Abelian $U(k)$ background gauge field. Charged fermions moving on ${\bf CP}^k$ can have non-Abelian degrees of freedom and can further couple to external non-Abelian gauge fields. In this paper we derive the bosonized action in the case of non-Abelian gauge interactions.  Although our explicit calculations involve the particular case of nonrelativistic fermions on ${\bf CP}^k$ whose LLL Hilbert space is well known \cite{KN3}, our method is quite general and applies to any manifold which admits a consistent fomrulation of the quantum Hall effect.

The semiclassical limit of the derived bosonic action, where $N \rightarrow \infty$ and the number of fermions becomes large, produces the collective description of the low-energy excitations of the quantum Hall droplet (Abelian and non-Abelian) in the presence of gauge interactions. In this limit the action separates into a boundary part and a bulk part. The boundary term is essentially described by a chiral, gauged Wess-Zumino-Witten type action, while the bulk term is a Chern-Simons type action in terms of the external gauge fields. These two contributions combine so that the total effective action is gauge invariant. The phenomenon of anomaly cancellation between the edge and bulk action is of course well known in two dimensions. There, one can easily integrate out the fermions and derive the bulk contribution which is an Abelian Chern-Simons action whose coefficient is given by the quantized Hall conductance. The Chern-Simons term defined on a space with boundary is not gauge invariant, the non-invariance given by a surface term. The gauge invariance is restored by the addition of a boundary action in terms of chiral massless fields, which describe the edge dynamics of a quantum Hall droplet \cite{wen, IKS}. In higher dimensions our bosonization method produces simultaneously the bulk and edge effective actions in a way that gauge invariance is automatically built in.

There have been several approaches to extending bosonization to higher dimensions \cite{boso1} -\cite{boso7}. Our approach is closer to the one followed by Das et al \cite{wadia} and Sakita \cite{sakita1, sakita2, shizuya} for LLL nonrelativistic fermions in two dimensions. Our noncommutative field theory action is essentially a $W_N$-gauge action in higher dimensions.

This paper is organized as follows. In section 2 we outline the bosonization method for nonrelativistic fermions in the LLL in the presence of gauge interactions in arbitrary dimensions. In the rest of the paper we apply the bosonization technique in the case of ${\bf CP}^k$ with a $U(k)$ background gauge field. In section 3 we review the LLL Hilbert space of charged fermions on ${\bf CP}^k$ with a $U(1)$ and $U(k)$ uniform background magnetic field. In section 4 we derive the corresponding star-product to $1 /n^2$ order, where $n$ is the strength of the $U(1)$ uniform magnetic field. In sections 5 and 6 we consider the large $N$ (equivalently large $n$) limit of the bosonic action and derive the edge and bulk effective action for the non-Abelian quantum Hall droplet. In section 7 we discuss the gauge invariance of the bosonic action and explicitly demonstrate the anomaly cancellation between the edge and bulk contributions. In section 8 we conclude with a brief summary and comments.

\section{General approach}

Here we present a general matrix formulation of the dynamics of noninteracting fermions in the lowest Landau level, which eventually leads to a bosonization approach in terms of a noncommutative field theory. 

Let $N$ denote the dimension of the one-particle Hilbert space corresponding to the states of the lowest Landau level, $K$ of which are occupied by fermions. The spin degree of freedom is neglected, so each state can be occupied by a single fermion. In the presence of a confining potential $\hat{V}$, the degeneracy of the LLL states is lifted and the fermions are localized around the minimum of the potential forming a droplet. The choice of the droplet we are considering is specified by a diagonal density matrix 
 $\hat {\rho}_0$ which is equal to 1 for occupied states and zero for unoccupied states. We can further consider $\hat{\rho}_0$ to be the density matrix characterizing the ground many-body state. The most general fluctuations which preserve the LLL condition and the number of occupied states are unitary transformations of $\hat {\rho}_0$, namely $\hat {\rho}_0 \rightarrow \hat{\rho}=\hat{U}  \hat {\rho}_0 \hat{U} ^ \dagger$, where  $\hat{U}$ is an $(N \times N)$ unitary matrix. The
action which determines ${\hat U}$ is given by 
\beq
S_{0}=  \int dt~ \Tr \left[ i  {\hat \rho}_0 { \hat U}^\dagger \del_t {\hat U}
~-~ {\hat \rho}_0 {\hat U}^\dagger {\hat{V}} {\hat U} \right]
\label{1}
\eeq
where $\hat {V}$ is the confining potential. We have used the fact that on the LLL the Hamiltonian is $\hat{V}$ up to an additive constant. $\hat{U}$ can be thought of as a collective variable describing all the possible excitations within the LLL. The equation of motion resulting from (\ref{1}) is the expected evolution equation for the density matrix $\hat{\rho}$, namely
\beq
i {{\partial \hat{\rho}} \over {\partial t}} = [ \hat{V} , \hat{\rho}]
\label{3}
\eeq

As it is well known by now in the context of noncommutative field theories, the action $S_{0}$ can also be written as \cite{KN2, KN3, KA}
\beq
S_{0}= N  \int d\mu dt~ \left[ i ({\rho}_0 *{  U}^\dagger  * \del_t { U})
~-~ ({ \rho}_0 *{U}^\dagger  * {{V}} * {U}) \right]
\label{4}
\eeq
where $d\mu$ is the volume measure of the space where QHE has been defined and $\rho_0,~U,~V$ are the symbols of the corresponding matrices on this space. In our notation the hatted expressions correspond to matrices and unhatted ones to the corresponding symbols, which are fields on the space where QHE is defined.  As we shall explain later, in the case where the LLL admits non-Abelian fermions coupled to a background gauge field in some representation $J'$ of dimension $N'$, the corresponding symbols are $(N' \times N')$ matrix valued functions and the action $S_0$ is written as
\beq
S_{0}= {N \over N'}  \int d\mu dt~ \tr~\left[ i ({\rho}_0 *{  U}^\dagger  * \del_t { U})
~-~ ({ \rho}_0 *{U}^\dagger  * {{V}} * {U}) \right]
\label{4a}
\eeq
Our notation is such that ``$\Tr$" indicates trace over the $N$-dimensional LLL Hilbert space while ``$\tr$" indicates trace over the $N'$-dimensional representation $J'$. In the case of Abelian fermions, which was the case studied in \cite{KN2, KA}, $N'=1$ and $\tr$ is absent as in (\ref{4}). An important point to emphasize here is that when we later consider the semiclassical limit of (\ref{4a}), $N \rightarrow \infty$, while $N'$ remains finite.

If $\Psi_m(\vec{x})$, $m=1,\cdots,N$,  represent the correctly normalized LLL wavefunctions, then the definition of the symbol corresponding to a $(N \times N)$ matrix $\hat{O}$, with matrix elements $O_{ml}$ is
\beq
O(\vec{x}, t) = { 1 \over N} \sum_{m,l} \Psi_m(\vec{x}) O_{ml}(t) \Psi^*_l(\vec{x})
\label{symb}
\eeq
The star product is defined as 
\beq
\big(\hat{O}_1\hat{O}_2\big)_{symbol} = O_1(\vec{x}, t) * O_2 (\vec{x}, t) 
\label{star}
\eeq
The action $S_{0}$ in (\ref{1}) or equivalently (\ref{4a}) provides an exact bosonization for the noninteracting fermion problem. The expression in (\ref{1}) does not depend on the particular space and its dimensionality or the Abelian or non-Abelian nature of the underlying fermionic system. This information is encoded in equation (\ref{4a}) in the definition of the symbol, the star product and the measure.

We now extend the bosonization approach for the underlying fermionic problem in the LLL in the presence of external, fluctuating gauge fields. We propose that gauge interactions (beyond the coupling to the strong uniform magnetic field which confines the system to the LLL) are described by a matrix action $S$ which is invariant under time dependent $U(N)$ rotations,
\beq
\hat{U} \rightarrow \hat{h} \hat{U} ~
\label{7}
\eeq
We consider $S$ to be the gauged version of $S_0$ in (\ref{1}), where $\del_t$ is now replaced by $\hat{D}_t= \del_t {\bf 1}+ i \hat{\A}$ and $\hat{\A}$ is a matrix gauge potential. In particular, 
\beq
S= \int dt~ \Tr \left[ i  \hat {\rho}_0  \hat {U}^\dagger ( \del_t + i \hat{\A} ) \hat {U}
~-~  \hat {\rho}_0 \hat {U}^\dagger \hat{V} \hat{U} \right]
\label{6}
\eeq
Invariance of this action under infinitesimal time dependent $U(N)$ rotations 
\beq
\delta\hat{U} = - i \hat{\l} \hat{U}
\label{7}
\eeq
implies the following transformation for the gauge potential $\hat{\A}$,
\beq
\delta \hat{\A}  =   \del _t {\hat{ \l }}-i [\hat{\l}, \hat{V}+\hat{ \A} ]
\label{8}
\eeq
where $\hat{h} = \exp (-i \hat{\lambda})$.
As before the action $S$ in (\ref{6}) can be written in terms of the corresponding symbols as
 \beq
S= {N \over N'}   \int dt~d\mu ~ \tr~\left[ i  \rho_0 * U^\dagger  * \del_t U
~-~ \rho_0 *U^\dagger * V *U- \rho_0 *U^\dagger * \A *U \right]
\label{10}
\eeq
where the ($N \times N$) matrices have been replaced by their symbols, matrix multiplication by the star product,  and $\Tr$ by ${N \over N'} \int d\mu~\tr $. The action (\ref{10}) is now invariant under the infinitesimal transformations
\beqar
\delta U & =& - i \l * U \nonumber \\
\delta \A (\vec{x}, t)  & = & \del _t \l (\vec{x}, t) - i \left( \l* (V+\A) - (V + \A) * \l \right) 
\label{11}
\eeqar
We shall refer to this as the $W_N$ gauge transformation, in analogy to the $W_\infty$ transformation appearing in the case of the planar two-dimensional QHE \cite{IKS, wadia}.

If the proposed action $S$ in (\ref{10}) is to be the bosonized action describing the coupling of the LLL fermionic system to a fluctuating external gauge field $A_{\mu}(\vec{x}, t)$, then ${\A}(\vec{x}, t)$ should be a function of $A_{\mu}(\vec{x},t)$. The dependence of ${\A}(\vec{x}, t)$ on $A_{\mu}(\vec{x}, t)$  is determined in the following way. Since $S$ is supposed to describe gauge interactions of the original system it has to be invariant under the usual gauge transformation
\beq
 \delta A_{\mu} = \del_{\mu} \Lambda
 \label{12}
 \eeq
in the case of Abelian gauge fields, or
 \beqar
 \delta A_{\mu} & = & \del_{\mu} \Lambda + i [\bar{A}_\mu + A_\mu , ~ \Lambda] 
 \label{13} \\
 \delta \bar{A}_{\mu} & = & 0 \nonumber 
 \eeqar
 in the case of non-Abelian gauge fields, where $\Lambda$ is the infinitesimal gauge parameter and $\bar{A}_\mu$ is a possible non-Abelian background field. The fact that $S$ is invariant under (\ref{11}) and (\ref{12}) (or (\ref{13})) implies that the transformation (\ref{11}) can be thought of as a nonlinear realization of the gauge transformation (\ref{12}) or (\ref{13}). This determines (up to gauge invariant terms) $\A$ as a function of $A_{\mu}$ and therefore the bosonized action of the LLL fermionic system in the presence of gauge interactions. Further, since ${\A}$ can be thought of as the time component of a noncommutative gauge field, the relation between ${\A}$ and the commutative gauge fields $A_\mu$ is essentially a Seiberg-Witten transformation \cite{seib1,seib2}.
 
 As we shall explicitly show in section 6, in the semi-classical limit, where $N \rightarrow \infty$ and the number of fermions is large, the ${\A}$-dependent part of the action produces a boundary term describing the coupling of the quantum Hall droplet to the external gauge field $A_\mu$, and also a purely $A_\mu$-dependent bulk term, which is a Chern-Simons like term. The bulk Chern-Simons term defined on a space with boundary is not gauge invariant. The gauge noninvariance is cancelled by the boundary term so that the total action is gauge invariant. At the large $N$ limit the transformations (\ref{11}) become the usual gauge transformations for matter and gauge fields so the gauge invariance of the action is automatically satisfied. 
 
 This approach, which is based on a matrix formulation, provides a very general way to construct the  bosonic action for the underlying LLL fermionic system in any space that admits a consistent formulation of QHE. The semiclassical limit of this action describes the gauge interactions of the quantum Hall droplet (Abelian or non-Abelian).
  
 The action $S_0$ in (\ref{4}) was also used  in the context of one dimensional free fermions and their relation to $c=1$ string theory \cite{wadia}. The bosonization approach of LLL fermions in the presence of gauge interactions extended to any dimension as outlined above, is an adaptation of a method used by Sakita \cite{sakita2} to derive the electromagnetic interactions of LLL spinless electrons in the two dimensional plane. 
 
 In \cite{KN2, KN3}, using a semi-classical expansion of $S_0$, we derived the edge dynamics of Abelian and non-Abelian quantum Hall droplets on higher dimensional ${\bf CP}^k$ spaces. In this case the fermionic density is a step function, constant over the phase volume occupied by the droplet and zero outside the droplet. We found that the action $S_0$ reduces to a higher dimensional generalization of a chiral Wess-Zumino-Witten type action describing the boundary excitations of the droplet. 
 In \cite{KA}  we further explored the semiclassical limit of $S$ in (\ref{10}) to derive the bulk and edge effective actions of the Abelian quantum Hall droplet in the presence of electromagnetic interactions. In this paper we extend our analysis to derive the edge and bulk dynamics of the non-Abelian quantum Hall droplet in the presence of gauge interactions.

 \section{ QHE on ${\bf CP}^k$}
 
 Here we shall briefly review the structure of the lowest Landau level and the emerging star product for ${\bf CP}^k$, which are the crucial ingredients in constructing the bosonic action (\ref{10}). We shall mainly follow the presentation in \cite{KN2, KN3} and in the review article \cite{KN4}. 

${\bf CP}^k$ is a $2k$-dimensional manifold parametrized by $k+1$ complex coordinates $v_a$, such that
\beq
\bar{v}_a v_a =1
\label{13a}
\eeq
with the identification $v_a \sim e^{i \theta} v_a$. One can further introduce local complex coordinates $z_{I}$, $I=1,\cdots, k$, by writing
\beqar
v_{I} & = & {z_{I} \over {\sqrt{1 + \bz \cdot z}}}~,~~~~~~~I =1, \cdots , k \nonumber \\
v_{k+1} & = & {1 \over {\sqrt{1 + \bz \cdot z}}} 
\label{complex}
\eeqar

The $U(1)$ background magnetic field (which leads to the Landau states) is introduced via a gauge potential
\beq
a = -i n  \bar{v} \cdot dv
\label{17}
\eeq
The corresponding $U(1)$ field strength is given by 
\beq
 d a = -in d\bar{v} \cdot dv = n \Omega
\label{18}
\eeq
where $\Omega$ is the K\"ahler two-form of ${\bf CP}^k$, which is obviously closed, and $n$ is an integer. Further $n=2 B R^2$, where $B$ is the constant background $U(1)$ magnetic field and $R$ is the radius of ${\bf CP}^k$.  

For $k=1$, since $S^2={\bf CP}^1$, the problem of charged fermions on ${\bf CP}^1$ with $U(1)$ background field was studied by Haldane several years ago \cite{haldane}. In this case the background gauge  field $a$ is that of a monopole of charge $n$ placed at the origin of $S^2$.

The  lowest Landau level wavefunctions for ${\bf CP}^k$ with $U(1)$ background field were derived in \cite{KN2}.  They are the coherent states for ${\bf CP}^k$.
\beqar
\Psi_m(\vec{x}) &=& \sqrt{N} \left[ {n! \over i_1! i_2! ...i_k!
(n-s)!}\right]^\half ~ {z_1^{i_1} z_2^{i_2}\cdots z_k^{i_k}\over
(1+\bz \cdot z )^{n \over 2}}~,~~~~~~~~~~~~m=1,\cdots, N  \nonumber\\
s &=& i_1 +i_2 + \cdots +i_k ,~~~0\le i_i \le n~,   0 \le s \le n \label{wav}
\eeqar
These wavefunctions form a symmetric, rank $n$ representation $J$ of $SU(k+1)$. The dimension of this representation is 
\beq
N={\rm dim} J = {{(n+k)!} \over {n! k!}} 
\label{dim}
\eeq
The set of states in (\ref{wav}) can also be interpreted as the Hilbert space of fuzzy ${\bf CP}^k$ \cite{bal}. QHE, therefore, provides a physical realization of fuzzy spaces \cite{KN4}, so our results can be relevant for certain fuzzy space analyses, going beyond the context of just QHE.

For ${\bf CP}^k$ one can also have uniform non-Abelian background fields. In this case the corresponding LLL wavefunctions are more involved. Using the fact that ${\bf CP}^k = SU(k+1)/U(k)$,
a group theoretic analysis was developed in \cite{KN3} which allowed a uniform treatment of the Abelian and non-Abelian case. Let $t_A$ denote the generators of $SU(k+1)$ as matrices in the fundamental representation, normalized so that $\tr (t_A t_B) = {1 \over 2} \delta_{AB}$. These generators are classified into three groups. The ones corresponding to the $SU(k)$ part of
$U(k) \subset SU(k+1)$ will be denoted by
$t_a$, $a =1, ~2, \cdots , ~ k^2 -1$ while the
generator for the $U(1)$
direction of the subgroup $U(k)$ will be denoted by 
$t_{k^2+2k}$. The $2k$ remaining generators of $SU(k+1)$ which are not in $U(k)$ are the coset generators, denoted by $t_\alpha$, $\alpha = 1,\cdots, 2k$. The coset generators can be further separated into the raising and lowering type $t_{\pm I}= t_{2I-1} \pm i t_{2I}, ~ I = 1, \cdots,k$. 

We can now use a $(k+1) \times (k+1)$ matrix $g$ in the fundamental representation of $SU(k+1)$ to parametrize ${\bf CP}^k$, by making the identification $g \sim gh$, where $h \in U(k)$. 
We can use the freedom of $h$ transformations to write $g$ as a function of the real coset coordinates $x^i$, $i=1,\cdots,2k$. The relation between the complex coordinates $z^I,~\bar{z}^I$ in (16) and $x^i$ is the usual one, $z^I = x^{2I-1} + i x^{2I},~I=1,\cdots,k$. We can write
\beq
g^{-1}dg = \big( -i E^{k^2+2k}_i t_{k^2+2k} -i E^{a}_i t_{a} -i E^{\alpha}_i t_{\alpha} \big)~ dx^i
\label{gdg}
\eeq
The $E^{\alpha}_i$ are the frame fields in terms of which the Cartan-Killing metric on ${\bf CP}^k$ is given by
\beq
ds^2 = g_{ij} dx^i dx^j = E^\alpha_i E^\alpha_j dx^i dx^j
\label{metric}
\eeq
The K\"ahler two-form on ${\bf CP}^k$ is likewise written as
\beqar
\Omega & = & -i \sqrt{{2k \over {k+1}}} \tr \left( t_{k^2 + 2k} ~ g^{-1}dg \wedge g^{-1}dg \right)
\nonumber
\\ & = & -{1 \over 4} \sqrt{{2k} \over {k+1}} f^{(k^2+2k)\alpha\beta} ~E^\alpha_i~ E^{\beta}_j ~dx^i \wedge dx^j \nonumber \\
& = & -{ 1 \over 4} \epsilon^{\alpha\beta} ~E^\alpha_i~ E^{\beta}_j ~dx^i \wedge dx^j~\equiv ~{1 \over 2} \Omega _{ij}~ dx^i
\wedge dx^j
\label{18d}
\eeqar
$f^{ABC}$ are the $SU(k+1)$ structure constants, where $[t_A,~t_B] = i f^{ABC} t_C$ and $\epsilon^{\alpha\beta} = 1$ if $\alpha= 2I-1, ~\beta= 2I, ~I=1, \cdots,k$.
The fields $E^{k^2+2k}_i$ and $E^a_i$ are related to the $U(1)$ and  $SU(k)$ background gauge fields on ${\bf CP}^k$. In particular the $U(1)$ field $a$ is given by
\beqar
a & = & i n \sqrt{{{2k} \over {k+1}}} \tr (t_{k^2+2k} g^{-1} dg ) = {n \over 2} \sqrt{{{2k} \over {k+1}}} E^{k^2+2k} \nonumber \\
& = & -i n~g^*_{a,k+1}dg_{a, k+1}
\label {Abar}
\eeqar
This agrees with (\ref{17}) if we identify $g_{a,k+1} = v_a$. We can similarly define an $SU(k)$ background field $\bar{A}^a_i$. Its normalization is chosen so that
\beq
\bar{A}^a \equiv E^a =  2i \tr (t^a g^{-1} dg) \label{NA}
\eeq
Notice that $\bar{A}^a$ in (\ref{NA}) does not depend on $n$, while the Abelian field $a$ in (\ref{Abar}) is proportional to $n$. 
The corresponding $U(1)$ and $SU(k)$ background field strengths are
\beqar
\del_i a_j - \del_j a_i & = & n \Omega_{ij} = - {n \over 2} \sqrt{{{2k} \over {k+1}}} f^{(k^2+2k)\alpha\beta} E^{\alpha}_iE^{\beta}_j  \nonumber \\
\bar{F}^{a}_{ij} & = & \del_i \bar{A}^a_j -\del_j \bar{A}^a_i + f^{abc} \bar{A}^b_i \bar{A}^c_j = - f^{a\alpha\beta} E^{\alpha}_iE^{\beta}_j
\label{barF}
\eeqar
We see from (\ref{barF}) that in the appropriate frame basis the background field strengths are constant, proportional to the $U(k)$ structure constants. It is in this sense that the field strengths in (\ref{barF}) correspond to uniform magnetic fields appropriate in defining QHE. 

We now define two sets of operators, $R_{A}$ and $L_A$ which perform right and left translations on an arbitrary element $g$ of $SU(k+1)$, as
\beq
R_A ~g = g~T_A~~~~~~~~~~~~~~~~~~~~L_A ~g = T_A~g
\label{RL}
\eeq
where $T_A$ are the $SU(k+1)$ generators in the representation to which $g$ belongs. 
The $U(1)$ gauge field in (\ref{Abar}) changes by a gauge transformation under a right $U(1)$ rotation of the form $g \rightarrow gh$ where $h \in U(1)$,  while it remains invariant under an $SU(k)$ right rotation. This implies that in the case where the fermions couple only to the Abelian gauge field $a$, the corresponding single particle wavefunctions have a fixed $U(1)_R$ charge and are singlets under $SU(k)$ right rotations \cite{KN2}. In particular the wavefunctions obey the condition
\beqar
 R_a ~\Psi_{m} &=& 0,~~~~~~~~~~~~~~~~a=1,\cdots,k^2-1  \nonumber \\
{R}_{k^2 +2k} ~\Psi_{m} &=& - {n k\over \sqrt{2 k
(k+1)}}~\Psi_{m} 
\label{9}
\eeqar
On the other hand, the non-Abelian gauge field $\bar{A}^a$ in  (\ref{NA}) is invariant under right $U(1)$ rotations but noninvariant under right $SU(k)$ rotations. So in the case where the fermions have non-Abelian degrees of freedom and couple to the full $U(k)$ background gauge field, the wavefunctions have the same fixed $U(1)_R$ charge as in (\ref{9}) but under right rotations transform as a particular $SU(k)$ representation $J'$ of dimension $N'= {\rm dim} J'$. In this case,
\beqar
 R_a ~\Psi_{m; a'}  &=& 
\Psi_{m;b'} ~ (T_a)_{b' a'} \nonumber \\ 
{R}_{k^2 +2k} ~\Psi_{m; a'} &=& - {n k \over \sqrt{2 k
(k+1)}}~\Psi_{m; a'} 
\label{9NA}
\eeqar
The indices $a' ,b'~=1,\cdots, N'$ label the states within the $SU(k)$ representation $J'$ and can be thought of as the internal degrees of freedom of the non-Abelian fermions coupled to the $U(k)$ background field. The matrices $T_a$ are the $SU(k)$ generators in the representation $J'$.

As we have explained in \cite{KN3, KN4}, the coset operators $R_{\alpha}$ correspond to covariant derivatives, while the $SU(k+1)$ operators ${L}_A$ correspond to magnetic translations. In particular, in the absence of a confining potential, the Hamiltonian is proportional to $R_{+I} R_{-I}$ up to additive constants, so the lowest Landau level condition is 
\beq
R_{-I} \Psi = 0
\label{LLLcondition}
\eeq

In both the Abelian and non-Abelian case the wavefunctions form an $SU(k+1)$ representation. A convenient basis to express them uniformly, is in terms of the Wigner ${\D}$-functions which are the matrices corresponding to the group elements in a particular representation $J$. Taking into account the proper normalization we have,
\beq
\Psi= \sqrt{N} {\D}^{(J)} _{L,R} (g) = \sqrt{N} ~\la J, l_A \vert~{\hat g}~ \vert J, r_A\ra
\label{wigner}
\eeq
where $l_A,~r_A$ indicate the two sets of quantum numbers specifying the states on which the generators act, for left and right actions respectively. 

In the Abelian case where the fermions couple to the $U(1)$ background gauge field $a$, the right state  $\vert J, r_A\ra$ in (\ref{wigner}) must be constrained by the condition (\ref{9}); we denote the right state $\vert J, r_A\ra$ by $\vert -n \ra$. Further, condition (\ref{LLLcondition}) implies that the state on the right  is the lowest weight state. As a result the LLL wavefunctions on ${\bf CP}^k$ with $U(1)$ background field form an irreducible $SU(k+1)$ representation $J$ which is symmetric, of rank $n$, and whose lowest weight state is an $SU(k)$ singlet. The dimension of the $J$ representation which defines the dimensionality of the LLL Hilbert space is given by (\ref{dim}).

In the non-Abelian case where the fermions couple to the $U(k)$ background field, the right states  $\vert J, r_A\ra$ in (\ref{wigner}) are constrained by the condition (\ref{9NA}); we now denote the right states $\vert J, r_A\ra$ by $ \vert a', -n \ra$.
Conditions (\ref{9NA}) and (\ref{LLLcondition}) imply that the LLL wavefunctions for ${\bf CP}^k$ with $U(k)$ background field form an irreducible $SU(k+1)$ representation $J$ whose lowest weight state is an $SU(k)$ representation $J'$. Since the $U(1)$ charge is fixed in terms of $n$, there are constraints on the type of allowed $J'$ $SU(k)$ representations \cite{KN3}. The dimension $N$ of the $SU(k+1)$ representation $J$ depends now on the particular $J'$ representation chosen, but for large $n$
\beq
N = {\rm dim} J \rightarrow {\rm dim} J' ~{n^k \over k!} = N' ~{n^k \over k!}
\label{35a}
\eeq
The LLL wavefunctions 
\beqar
\Psi_{m; a'} (g) &=& \sqrt{N} ~\la J, l_A \vert ~{\hat g} ~\vert a', - n\ra\nonumber\\ 
&\equiv& \sqrt{N}~ {\cal D}_{m;a'}(g)
\label{35}
\eeqar
are properly normalized by virtue of the orthogonality theorem
\beq
\int d\mu (g) ~\D^*_{m;a'} (g)~\D_{l;b '}
(g) ~=~ {\delta_{ml}\delta_{a' b '}\over N}
\label{orthon}
\eeq

\section{Star product for ${\bf CP}^k$ with $U(k)$ background gauge field}

As we have discussed in \cite{KN2, KN3} the symbol corresponding to a $(N \times N)$ matrix $\hat{X}$, with matrix elements $X_{ml}$, acting on the Hilbert space of the LLL is defined by
\beqar
X_{a'b'} (\vec{x}, t) &  =  & {1 \over N} \sum_{ml}{\Psi}_{m;a'}(\vec{x}) ~X_{ml} (t)  {\Psi}^*_{l;b'}(\vec{x}) \nonumber \\
& = & \sum_{ml}{\cal D}_{m;a'}(g) ~X_{ml} {\cal
D}^*_{l;b'}(g) \nonumber \\
& = & \la b',-n \vert {g}^{\dagger} X^T {g} \vert a';-n \ra
\label{symbolNA}
\eeqar
In the non-Abelian case the symbol is a $(N' \times N')$ matrix valued function, while in the Abelian case where $J'$ is the singlet representation, the symbol is just a function on ${\bf CP}^k$. With this definition
\beq
\Tr {\hat X}  =  {N \over N'} \sum_{a'} \int d\mu (g) ~X_{a'a'} (g)
\label{}
\eeq
The star product is defined in terms of the symbol corresponding to the product of two matrices $\hat{X}$ and $\hat{Y}$,
\beqar
(\hat{X}\hat{Y})_{a'b'} & = & X_{a'c'} *  Y_{c'b'} \nonumber \\                                  
& = & \sum_{mrl} {\cal D}_{m;a'}(g) ~X_{mr}Y_{rl}
{\cal D}^{*}_{l;b'}(g) \nonumber \\
& = & \la b',-n \vert {g}^\dagger Y^T X^T {g} \vert a',-n \ra \nonumber \\
& = & \la b',-n \vert {g}^\dagger Y^T ~ {\bf 1}~X^T {g} \vert a',-n \ra
\label{star}
\eeqar
In order to calculate the star product we need to reexpress the unit matrix ${\bf 1}$ in (\ref{star}), where ${\bf 1} = \sum_m \vert m \ra \la m \vert$, and $\vert m \ra$ are all the states in the $J$ representation, in terms of the lowest weight states $\vert a',-n \ra$.  In the case of a $U(1)$ background field the star product, following this method, was derived in \cite{KN2}. We found
\beqar
X*Y &=& \sum_s (-1)^s \left[ {(n-s)! \over n! s!}\right]
\sum_{i_1+i_2+\cdots +i_k=s}^n~
{s! \over i_1! i_2! \cdots i_k!}~{ R}_{-1}^{i_1} { R}_{-2}^{i_2}
\cdots { R}_{-k}^{i_k} X \nonumber\\
&&\hskip 2.5in\times~
{ R}_{+1}^{i_1} {R}_{+2}^{i_2}\cdots { R}_{+k}^{i_k}Y
\label{25a}
\eeqar
Expression (\ref{25a}) can be thought of as a series expansion in $1/n$.

In the case of the $U(k)$ background field the calculation of the star product is more involved. In \cite{KN3} we calculated it to order $1/n$; here we shall extend the calculation to order $1/n^2$, which is sufficient for the derivation of the effective action $S$ in the semiclassical limit $N \rightarrow \infty$ (or equivalently $n \rightarrow \infty$).

 An arbitrary state in the $SU(k+1)$ representation $J$ can be written as
 \beq
 \vert \Psi \ra = \sum_{a'} C_{a'} \vert a', -n \ra + \sum_{a'I} C_{a'}^{I} T_{+I} \vert a' , -n \ra + \sum_{a'IJ}C_{a'} ^{\{IJ\}} T_{+I}T_{+J} \vert a',-n \ra + \cdots 
 \label{complete}
 \eeq
 where $\vert a',-n \ra$ are the lowest weight states within the representation $J$, $T_{+I}~I=1,\cdots,k$ are the corresponding raising generators and $C_{a'}$ are coefficients to be determined. Using the commutation relation
\beq
[ T_{-I} , T_{+J} ] = - \sqrt{ {2(k+1) \over k}}~ T_{k^2+2k}~ \delta_{IJ}
+i f^{a\bar{I}J}~ T_a\label{15b}
\eeq
and the fact that $\sqrt{2 k (k+1)}~T_{k^2+2k} T_{+I} \vert a',-n \ra =( -nk +k+1)T_{+I} \vert a',-n \ra$, we find the following relations for the coefficients $C$ :
\beqar
 \la b', -n \vert \Psi \ra & = & C_{b'} \nonumber \\
\la b', -n \vert T_{-J} \Psi \ra & = & \sum_{a',I}  C_{a'}^{I} \left[ n \delta_{IJ} \delta_{a'b'} +i f^{a\bar{J}I} \la b',-n \vert T_a \vert a' ,-n \ra \right] + \cdots  \nonumber \\
\la b', -n \vert T_{-K}T_{-L} \Psi \ra & = & \sum_{a',I,J} C_{a'}^{\{IJ\}}  \left[ n^2 (\delta_{LI} \delta_{KJ}  +\delta_{KI} \delta_{LJ} ) \right] \delta_{a'b'} + \cdots 
\label{C-1}
\eeqar
where $\cdots$ above denotes terms of lower order in $n$.
Inverting the last two expressions in (\ref{C-1}) we find
\beqar
C_{a'}^{I} & = &  \sum_{J}\left[  {1 \over n} \delta_{a'b'} \delta_{IJ} - {i \over n^2} f^{a\bar{I}J} (T_a)_{a'b'} \right] \la b',-n \vert T_{-J} \Psi \ra  + ~{\cal{O}} ({1 \over n^3}) \nonumber \\
C_{a'}^{\{IJ\}} & = &  {1 \over {4 n^2}} \sum_{KL} \delta_{a'b'} \big( \delta_{IK}\delta_{JL} + \delta_{IL}\delta_{JK} \big)   \la b',-n \vert T_{-K}T_{-L}  \Psi \ra  + ~{\cal{O}} ({1 \over n^3}) 
\label{C}
\eeqar
Combining (\ref{complete}), (\ref{C-1}) and (\ref{C}) we find
\beqar
{\bf 1} & = &\sum_{a'} \vert a', -n \ra \la a',-n \vert ~+~ 
\sum_{a'b' IJ} T_{+I} \vert a',-n \ra \left[ {1 \over n} \delta_{a'b'} \delta_{IJ} - {i \over n^2} f^{a\bar{I}J} (T_a)_{a'b'} \right]
~\la b' ,-n \vert T_{-J} \nonumber \\
&+&~{1 \over {2n^2}} ~T_{+I}T_{+J} \vert a',-n \ra \la a',-n \vert T_{-I}T_{-J} ~+ ~  {\cal{O}} ({1 \over n^3})
\label{42}
\eeqar 
Inserting (\ref{42}) in (\ref{star}) and using the definition of the symbol as in (\ref{symbolNA}) we find
\beqar
X*Y & = & XY - { 1 \over n} R_{-I}X R_{+I}Y + {i \over n^2} R_{-J} X f^{a \bar{I}J} (T_a)^{T} R_{+I} Y \nonumber \\
& + & {1 \over {2n^2}} R_{-I}R_{-J} X R_{+I} R_{+J} Y~+~ {\cal{O}}({1 \over n^3})
\label{43}
\eeqar
Using (\ref{gdg}) and (\ref{RL}) we can further express $R_{\alpha}$ as differential operators in the following way.
\beqar
R_\alpha {g} & = & i (E^{-1})^i_\alpha \big( \del_i {g} + i {g} ~E_i^{k^2+2k} T_{k^2+2k} +i {g}~ E_i^a T_a \big) \equiv i (E^{-1})^i_\alpha D_i {g} \nonumber \\
R_\alpha {g}^\dagger & = & i (E^{-1})^i_\alpha \big( \del_i {g}^\dagger - i  E_i^{k^2+2k} T_{k^2+2k} ~{g}^\dagger -i  E_i^a T_a~ {g}^\dagger \big) \equiv i (E^{-1})^i_\alpha D_i {g}^\dagger 
\label{44}
\eeqar
where $T$'s are the $U(k)$ generators in the particular representation ${g}$ belongs to. Using (\ref{44}) and the definition (\ref{symbolNA}) for the symbol, we find that the action of the right operator $R_\alpha$ on a symbol is
\beqar
R_\alpha X_{a'b'} & = & i (E^{-1})^i_\alpha ( D_i X)_{a'b'} \nonumber \\
D_i X & = & \del_i X + i [\bar{A}_i ,~X],~~~~~~~~~~~\bar{A}_i = \bar{A}^a_i (T_a)^T= E_i^a (T_a)^T
\label{44a}
\eeqar
where $\bar{A}$ is the $SU(k)$ background gauge field in the $J'$ representation. Notice that the $U(1)$ part of the gauge field does not contribute in (\ref{44}). \footnote{The particular definition of the symbol in (\ref{symbolNA}) implies that the gauging is done in terms of the transpose matrices $(T_a)^T$as in (\ref{44}). This was incorrectly stated in equation (53) of reference \cite{KN3}, where $T_a$ there should be replaced by $-(T_a)^T$.}

Similarly the action of two right operators $R_\alpha$ on a symbol produces the following expression
\beqar
R_\alpha R_\beta X_{a'b'} & = & - (E^{-1}) ^i_\alpha D_i \big((E^{-1})^j_\beta X_{a'b'}\big) \nonumber \\
& = & - (E^{-1})^i_\alpha (E^{-1})^j_\beta {\cal{D}}_i D_j X_{a'b'} 
\label{45}
\eeqar
where ${\cal{D}}_i$ is the properly defined covariant derivative for a curved space such as ${\bf CP}^k$, namely
\beqar
{\cal{D}}_iD_j X & \equiv & D_iD_j X - \Gamma^l_{ij} D_l X \nonumber \\
D_i E^\alpha_j & = & \del_i E_j^{\alpha} + f^{\alpha A \beta} E_i^A E_j^\beta ~ = ~ \Gamma_{ij}^l E_l^\alpha
\label{46}
\eeqar
where $A$ in $f^{\alpha A \beta}$ is a $U(k)$ index (both $U(1)$ and $SU(k)$) and $\Gamma_{ij}^l$ is the Christoffel symbol for ${\bf CP}^k$. 

Combining expressions (\ref{44}) and (\ref{45}) we can rewrite the star-product in (\ref{43}) in terms of covariant derivatives and real coordinates (instead of complex) as
\beqar
X*Y & = & XY + {1 \over n} P^{ij} D_i X D_j Y - {i \over n^2} P^{il} P^{kj} D_i X \bar{F}_{lk} D_j Y \nonumber \\
& + & { 1 \over {2 n^2}} P^{ik} P^{jl} {\cal{D}}_iD_j X {\cal{D}}_kD_l Y + {\cal{O}}({1 \over n^3})
\label{47}
\eeqar
where $\bar{F}_{lk} = \bar{F}^a_{lk} (T_a)^T$ and
\beq
P^{ij} = g^{ij} + {i \over 2} (\Omega^{-1})^{ij} 
\label{48}
\eeq
In deriving (\ref{47}) the following expressions for $g^{ij} $ and $(\Omega^{-1})^{ij}$ were used:
\beqar
g^{ij} & = & E^i_{\alpha} E^j_\alpha \nonumber \\
(\Omega^{-1})^{ij} & = & 2 ~\epsilon^{\alpha\beta}~ E^i_{\alpha} E^j_\beta 
\label{48a}
\eeqar
Using (\ref{47}) we can now write down the symbol for the commutator to order $1/ n^2$, 
\beqar
[X,~Y]_* & \equiv & X*Y - Y*X = [X,~Y] + { 1 \over n} P^{ij} (D_i X D_j Y - D_i Y D_j X) \nonumber \\
& & - {i \over n^2} P^{il} P^{kj} (D_i X \bar{F}_{lk} D_j Y - D_i Y \bar{F}_{lk} D_j X) \nonumber \\
&& + { 1 \over {2 n^2}} P^{ik} P^{jl} ( {\cal{D}}_iD_j X {\cal{D}}_kD_l Y - {\cal{D}}_iD_j Y {\cal{D}}_kD_l X) +~{\cal O}( {1 \over n^3})
\label{50}
\eeqar
Equations (\ref{47}) and (\ref{50}) are valid for both the Abelian and non-Abelian case. In the Abelian case, they simplify by taking $ X,~Y $ to be commuting functions, $\bar{F}_{lk} \rightarrow 0$, $D_i X \rightarrow \del_iX$ and ${\cal{D}}_iD_j X \rightarrow  \del_i\del_j X -\Gamma_{ij}^l \del_l X$.

\section{Calculation of ${\A}$}

As we explained in section 2, our approach in deriving the bosonized action $S$ expressing the dynamics of the LLL fermions on ${\bf CP}^k$ in the presence of gauge interactions results in calculating $\A$ as a function of the fluctuating gauge fields $A_\mu$ via the $W_N$ transformation (\ref{11}) and the fact that this is induced by the gauge transformation (\ref{13}).
Using (\ref{11}) and (\ref{50}) we find 
\beqar
\delta \A & = & \del_t \lambda - i[\lambda,~V+\A] - { i \over n} P^{ij} \left( D_i \lambda D_j (V + \A) - D_i (V + \A) D_j \lambda \right)  \nonumber \\
& & - {1 \over n^2} P^{il} P^{kj} \left( D_i \lambda \bar{F}_{lk} D_j V - D_i V \bar{F}_{lk} D_j \lambda\right)  \nonumber \\
&& - { i \over {2 n^2}} P^{ik} P^{jl} \left( {\cal{D}}_iD_j \lambda {\cal{D}}_kD_l V - {\cal{D}}_iD_j V {\cal{D}}_kD_l \lambda\right)
\label{51}
\eeqar

Before we attempt to explicitly solve for $\A$ as a function of $A_\mu$ we need to discuss the scaling of various quantities. All expressions so far (including the measure $d\mu$, $g^{ij}$, $(\Omega^{-1})^{ij}$, etc.) have been written in terms of the dimensionless coordinates $x_i = \tilde{x}_i/R$, where $R$ is the radius of ${\bf CP}^k$ and $\tilde{x}$ are the dimensionful coordinates. The calculation of the star-product (\ref{47}) involves a series expansion in terms of $1/n$, where $n=2 B R^2$ and $B$ is the constant $U(1)$ magnetic field. Rewriting our expressions in terms of the dimensionful parameters $\tilde{x}_i$, one can easily see that the expansion in  $1/n$ becomes an expansion in $1/B$. We further assume that the energy scale of the fluctuating gauge field $A_\mu$ and therefore $\A$ is much smaller than $B$ so that the restriction to LLL is justified. The scale of the confining potential $V$ is set by the magnetic field $B$ ($\sim n$ in terms of dimensionless variables) \cite{KN2, KN3}. A convenient choice for the confining matrix potential $\hat{V}$ is such that the ground state density  $\rho_0(\vec{x})$ corresponds to a spherical droplet. This is the case when all the $SU(k)$ multiplets of the $J$ representation 
upto a fixed hypercharge are completely filled, starting from
the lowest. A particular, simple choice for such a potential is the one used in \cite{KN3}, 
\beq
\hat{V} = \sqrt{2k \over k+1}~\omega ~\left( { T}_{k^2+2k} + {nk \over
\sqrt{2k(k+1)}}
\right) \label{52}
\eeq
where $\omega$ is a constant. (The potential does not have to be exactly of this form;
any potential with the same qualitative features will do.)  The particular expression (\ref{52}) is such
that
\beq
\la s \vert \hat{V} \vert s \ra = \omega s 
\label{53}
\eeq
where $\vert s \ra$ denotes an $SU(k)$ multiplet of hypercharge $-nk +sk + s$, namely
$\sqrt{2k(k+1)}{T}_{k^2+2k} \vert s \ra = (-nk +sk + s) \vert s \ra$. The symbol for (\ref{52}) was calculated in \cite{KN3} to be
\beqar
V_{a'b'} &=& \la b',-n \vert {g}^\dagger {V}^T  {g} \vert a',-n\ra \nonumber \\ 
& = & \omega n {{\bar{z} \cdot z} \over {1 + \bar{z} \cdot z}} \delta_{a' b'} ~+~S_{k^2+2k,a} (T_a)_{b' a'}\label{54}
\eeqar
where
\beq
S_{k^2+2k,a} = 2 \tr ({g}^\dagger t_{k^2+2k} ~{g}~ t_{a})\label{55}
\eeq
The important point is that the first term in (\ref{54}) is diagonal and of order $n$ in terms of the dimensionless variables $z$, while the second non-diagonal term is of order $n^0$. So in analyzing (\ref{51}) we can absorb the order $n^0$ term of the confining potential in the definition of $\A$ and treat separately the diagonal term of order $n$ as  $V$. Using then that $V_{a'b'}$ is a commuting diagonal matrix of the form $V_{a'b'} = \delta_{a'b'} V(r)$, where $r^2=\bar{z}\cdot z$, the expression (\ref{51}) can be further simplified as
\beqar
\delta \A & = & \del_t \lambda - i[\lambda,~\A] + {1 \over n} (\Omega^{-1})^{ij} D_i \lambda \del_j V - { i \over n} P^{ij} \left(D_i \lambda D_j \A - D_i \A D_j \lambda \right) \nonumber \\
& & - {1 \over n^2} P^{il} P^{kj} \left(D_i \lambda \bar{F}_{lk} \del_j V - \bar{F}_{lk} D_j \lambda \del_i V \right) \nonumber \\
&& + { 1 \over {2 n^2}} \left[(\Omega^{-1})^{ik} g^{jl} + g^{ik}  (\Omega^{-1})^{jl} \right] {\cal{D}}_iD_j \lambda 
\nabla_k \del_lV
\label{56}
\eeqar
where
\beq
\nabla_k \del_l V = \del_k \del_l V - \Gamma^n_{kl} \del_n V
\label{56a}
\eeq
A consistent solution for $\A$ as a function of $A_\mu$, $\A = f(A_\mu)$, is such that 
\beqar
\delta \A & = & f(\delta A_\mu) \nonumber \\
\delta A_\mu & = & \del_\mu \Lambda + i [\bar{A}_\mu + A_\mu, \Lambda]= D_\mu \Lambda + i [A_\mu, \Lambda]  \label {57} \\
\delta \bar{A}_\mu & = & 0 \nonumber
\eeqar
In the absence of a confining potential, $V=0$, the solution to (\ref{56}) is
\beqar
\A^{V=0} & = & A_0 - { i \over 2n} P^{ij} (A_iX_j - X_i A_j) \nonumber \\
X_i & = & 2 D_i A_0 - \del_0 A_i + i [A_i,~A_0]  \label{58} 
\eeqar
and
\beq
\lambda = \Lambda + { i \over {2n}} P^{ij} (D_i \Lambda A_j - A_i D_j \Lambda) \label{59}
\eeq
Writing
\beq
\A = \A^{V=0} + \A^{V}
\label{60}
\eeq
where $\A^V$ is the $V$-dependent part of $\A$ and using (\ref{56}) and (\ref{58}) we find the following relation for $\A^{V}$
\beqar
\delta \A^{V} & = & u^i D_i \lambda - i [\lambda,~\A^V]  - { i \over n} P^{ij} \left(D_i \lambda D_j \A^V - D_i \A^V D_j \lambda \right) \nonumber \\
& & - {i \over n} \left(P^{il} D_i \lambda \bar{F}_{lk} - P^{li} \bar{F}_{lk} D_i \lambda \right) u^k \nonumber \\
&& + { 1 \over {2 n^2}} \left[(\Omega^{-1})^{ik} g^{jl} + g^{ik}  (\Omega^{-1})^{jl}\right]  {\cal{D}}_iD_j \lambda 
\nabla_k \del_l V
\label{61}
\eeqar
where 
\beq
u^i = { 1 \over n} (\Omega^{-1})^{ij} \del_j V
\label{62}
\eeq
The quantity $u^i$, which will be extensively used from now on, is essentially the phase space velocity, if we think of the LLL as the phase space of a lower dimensional system, with symplectic structure $n \Omega $ and Hamiltonian $V$. 

In deriving (\ref{61}) the following relations between $g^{ij},~ (\Omega^{-1})^{ij}$ in (\ref{48a}) and the background field strength $\bar{F}^a_{ij}$ in (\ref{barF}) were used
\beqar
(\Omega^{-1})^{ij} \bar{F}_{ij}^a & = & 0 \nonumber \\
g^{ki} g^{lj} \bar{F}^{a}_{kl} & = & { 1\over 4} (\Omega^{-1})^{ki} (\Omega^{-1})^{lj} \bar{F}^{a}_{kl} \nonumber \\
g^{ki} (\Omega^{-1})^{lj} \bar{F}^{a}_{kl} & = & - (\Omega^{-1})^{ki} g^{lj}  \bar{F}^{a}_{kl} 
\label{63}
\eeqar
We eventually find that the solution for $\A^V$ is 
\beqar
\A^V & = & u^i A_i -{ i \over {2n}} P^{ij} (A_i A_k \del_j u^k - A_k A_j \del_i u^k) \nonumber \\
&& -{i \over {2n}} P^{ij} [A_i (X_{jk} + 2 \bar{F}_{jk}) - (X_{ik} + 2 \bar{F}_{ik}) A_j] u^k \nonumber \\
&& + { 1 \over {2n^2}} [(\Omega^{-1})^{ik} g^{jl} + g^{ik}  (\Omega^{-1})^{jl}] {\cal{D}}_i A_j \nabla_k\del_l V
\label{64}
\eeqar
where
\beq
X_{jk} = 2 D_j A_k - D_k A_j + i [ A_j,~A_k]
\label{65}
\eeq
and $\lambda$ is as in (\ref{59}), independent of $V$.
Combining (\ref{58}), (\ref{60}), (\ref{64})  and (\ref{65}) we arrive at the following expression for $\A$:\beqar
\A & = & A_0 -{i \over 2n} g^{ij} \left[ A_i,~2 D_i A_0 - \del_0 A_i + i [A_i,~A_0] \right] + { 1 \over {4n}} (\Omega^{-1})^{ij} \{A_i, 2 D_j A_0 - \del_0 A_j + i [A_j,~A_0] \} \nonumber \\
&& + u^i A_i -{i \over {2n}} g^{ij} \left[A_i,~A_k \right] \del_j u^k + {1 \over {4n}} (\Omega^{-1})^{ij} \{A_i,~A_k \} \del_j u^k \nonumber \\
&&-{i \over {2n}} g^{ij} \left[ A_i,~  2 D_j A_k - D_k A_j + i [ A_j,~A_k] ~ + 2 \bar{F}_{jk} ~\right] u^k \nonumber \\
&& + {1 \over {4n}} (\Omega^{-1})^{ij} \{ A_i,~  2 D_j A_k - D_k A_j + i [ A_j,~A_k] + 2 \bar{F}_{jk} ~ \} u^k \nonumber \\
&& + {1 \over {2n^2}} g^{ik} (\Omega^{-1})^{jl} \left( {\cal{D}}_i A_j + {\cal{D}}_j A_i \right) \nabla_k\del_l V
\label{66}
\eeqar
where $[~]$ indicate commutators and $\{~\}$ anticommutators.
The external fluctuating gauge field $A_\mu$ in the above expression, contains both the Abelian $U(1)$ and non-Abelian $SU(k)$ components, namely $A_\mu = A_\mu^a (T_a)^T + A_\mu^{k^2+2k} (T_{k^2+2k})^T$. 
In the Abelian case where the fermions interact only with the $U(1)$ gauge field, the symbols are commuting functions, so the commutator terms in (\ref{66}) vanish. The result then agrees with the one found in \cite{KA}. \footnote{The last term in (\ref{66}) was neglected in \cite{KA}.} In terms of the dimensionful quantities $\tilde{x}= R x$ , $\tilde{D}= D/R,~\tilde{A}=A/R,~\tilde{V} \sim B$, $\A$ can be written as a series expansion in $1/B$.  The terms we have kept in (\ref{66}) account for all terms of order $B^0$ and $1/B$.

$\A$ being the symbol of the time component of the matrix gauge potential, expression (\ref{66}) along with (\ref{59}) can be thought of as the Seiberg-Witten map \cite{seib1,seib2} for a curved manifold in the presence of non-Abelian background gauge fields.

As should be clear from (\ref{57}), expression (\ref{66}) is only determined up to gauge invariant terms whose coefficients are not constrained by the $W_N$ transformation (12) and the requirement that it is induced via the gauge transformation (14). We shall refer to the solution (\ref{66}) as the ``minimal" solution. As we shall see later, this produces the minimal gauge coupling for the chiral field describing the edge excitations of the quantum Hall droplet, similarly to the case of the Abelian droplet \cite{KA}.

\section{Edge and bulk actions}

The fully bosonized action expressing the gauge interactions of nonrelativistic fermions with $U(k)$ degrees of freedom with a fluctuating gauge field $A_\mu$ in the lowest Landau level is given by
\beq
S= {N \over N'}   \int dt~d\mu ~ \tr  \left[ i  \rho_0 * U^\dagger  * \del_t U
~-~ \rho_0 *U^\dagger * V *U- \rho_0 *U^\dagger * \A *U \right]
\label{67}
\eeq
where $\A$ is given in (\ref{66}). 

For the case of a confining potential $\hat{V}$ with an $SU(k)$ symmetry, as discussed in the previous section, the fermionic many-body ground state is formed by filling up a certain number of complete $SU(k)$ representations, starting with the singlet, the fundamental, rank two and so on, up to, let us say rank $M$ symmetric representation. We found in \cite{KN3} that in the large $N$, large $M$ limit, where $N \gg M$
\beqar
(\rho_0)_{a'b'}  & = & \rho_0(r^2) \delta_{a'b'} \nonumber \\
\rho_0(r^2) & = & \Theta \Big(1 - {{n r^2} \over M}\Big) = \Theta \Big(1 - {{R^2 r^2} \over {R_D^2}}\Big)
\label{step}
\eeqar
where $\Theta$ is the step function, and $R_D$ is the radius of the droplet, $R_D^2 = {M \over {2B}}$.  Equation (\ref{step}) defines the so-caled droplet approximation for the fermionic density, and it is at this limit we want to evaluate the action $S$ and identify the edge and bulk effective actions. 

As we mentioned earlier, the integrand in (\ref{66}) can be thought of as an expansion in $1/B$ if we write our expressions in terms of the dimensionful coordinates $\tilde{x}$. Similarly, using (\ref{35a}), the prefactor $(N/N') d\mu \rightarrow [n^k /(k! R^{2k})] d \tilde{\mu} = (2B)^k/k!~d\tilde{\mu}$, where $d\tilde{\mu}$ is the measure of the space in terms of the dimensionful coordinates. For convenience  we will continue the evaluation of  the edge and bulk effective actions in terms of the dimensionless coordinates, keeping in mind, though, that the $1/n$ expansion can always be converted to a $1/B$ expansion with the appropriate overall prefactor to  correctly accomodate the volume of the droplet.  

\subsection{Calculation of $S_0$}

In the absence of gauge interactions the semiclassical limit of $S_0$ was derived in \cite{KN3}. Here we give a brief review of this calculation.
\beq
S_0 = {N \over N'}  \int dt d\mu ~\tr \left( \rho_0 * U^\dagger * i \del_t U ~-~\rho_0 * U^\dagger * V * U \right)
\label{67a}
\eeq
$U$ can be expressed in terms of the hermitian 
 $(N' \times N')$ matrix valued field $\Phi$, which is the symbol corresponding to $\hat{\Phi}$ in $\hat{U} = e^{i \hat{\Phi}}$. We found that to leading order in $1/n$
 \beqar
U& = & G ~-~ {i\over n} G~F + \cdots \nonumber \\
G & = &  e^{i \Phi} \nonumber \\
F& = & -P^{ij} \int_0^1 d\alpha ~ e^{-i\alpha \Phi} ~D_i \Phi~
D_j (e^{i\alpha\Phi })
\label{23a}
\eeqar
Similarly
\beq
U^\dagger = G^\dagger ~+~{i\over n} F^\dagger G^\dagger ~+\cdots
\label{24a}
\eeq
and further
\beq
F- F^\dagger = i P^{ij} G^\dagger D_i G ~ G^\dagger D_j G 
\label{30a}
\eeq
Using (\ref{23a})-(\ref{30a}) one can show that the action $S_0$ can be written in terms of the unitary $(N' \times N')$ matrix valued field $G$, and in the large $n$-limit
\beqar
S_0 &=& {N \over {2nN'}} \int  dt d\mu ~\del_i \rho_0 (\Omega^{-1})^{ij} 
\tr \left[ \left( G^{\dagger} {\dot G} + u^k ~G^{\dagger} D_k G \right)  
G^{\dagger}D_j G  \right] \nonumber\\
&&-{N \over {2nN'}} \int  dt d\mu ~ \rho_0 (\Omega^{-1})^{ij} 
 \tr \left[ G^{\dagger}{\dot G}~ 
 G^{\dagger}D_iG ~G^{\dagger}D_jG \right]
\label{68}
\eeqar
Expanding out the covariant derivatives we find that the last term in (\ref{68}) can be written as \cite{KN3}
\beqar
&& \int  dt d\mu ~  (\Omega^{-1})^{ij}  \rho_0 
 \tr \left( G^{\dagger}{\dot G}~ 
 G^{\dagger}D_iG ~G^{\dagger}D_jG \right) \nonumber \\
 & = & \int  dt d\mu ~  (\Omega^{-1})^{ij}  \rho_0 \tr \left\{
   \left( G^{\dagger}{\dot G}~ 
 G^{\dagger}\del_iG ~G^{\dagger}\del_jG \right) 
 - \del_i \left[ ~i \left({\dot G} G^\dagger + G^\dagger {\dot G} \right) \bar{A}_j \right] \right\}
 \label{68a}
 \eeqar

Since $\rho_0(r^2)$ is a step function as in (\ref{step}), its derivative $\del_i \rho_0$ produces a delta function with support at the boundary 
of the droplet, namely
\beqar
\del_i \rho_0 & = & 2 r \hat{x}_i {{\del \rho_0} \over {\del r^2}} \nonumber \\
{{\del \rho_0} \over {\del r^2}} & = &  -{n \over M} \delta \left( 1 - {{nr^2} \over M} \right) = -{R^2 \over {R^2_D}} ~\d \left( 1 - {{R^2 r^2} \over {R^2_D}}
\right)
\label{69}
\eeqar
where $\hat{x}_i$ is the radial unit vector normal to the boundary of the droplet.
Further using the identities\footnote{The normalization used is such that $\int d\mu =1$.}
\beqar
&& d \mu = \epsilon^{i_1j_1i_2j_2\cdots i_kj_k}\Omega_{i_1j_1} \cdots \Omega_{i_{k}j_{k}} ~{d^{2k} x \over {(4 \pi)^k}} = k! \sqrt{\det \Omega}~ {d^{2k} x \over {(2 \pi)^k}}\nonumber \\
&& d\mu~(\Omega^{-1})^{ij}= -{2k}  ~\epsilon^{iji_2j_2 \cdots i_{k}j_{k}} \Omega_{i_2j_2} \cdots \Omega_{i_{k}j_{k}} ~{d^{2k} x \over {(4 \pi)^k}}
\label{dmuo}
\eeqar
we can rewrite $S_0$ in the following form:
\beqar
S_0
&&= -{N \over {2nN'}} \int  dt d\mu ~{{\del \rho_0} \over \del r^2} ~
 \tr \left[ \left( G^{\dagger} {\dot G} + \omega ~G^{\dagger} D_{\Omega} G \right)  
G^{\dagger}D_{\Omega} G  \right] \nonumber \\
&&+ {{Nk} \over { 4 \pi nN'}} \int \rho_0 \left[ -d \left( i \bar{A} dG G^{\dagger} + i \bar{A} G^{\dagger}dG \right) + {1 \over 3} \left( G^{\dagger}dG \right)^3 \right] \wedge \left( {\Omega \over {2 \pi}}\right)^{k-1} 
\label{70}
\eeqar
where $\omega = {1 \over n} {\del V \over {\del r^2}} \vert _{\rm boundary}
$ and 
\beq
D_{\Omega} = - (\Omega^{-1})^{ij} 2 r \hat{x}_i D_j
\label{71}
\eeq
$D_{\Omega}$ is the component of the covariant derivative $D$ perpendicular to the radial direction, along a special tangential direction on the droplet boundary. The action $S_0$ in (\ref{70}) is a higher dimensional generalization of a chiral,  Wess-Zumino-Witten action, vectorially gauged \cite{nepomechie}
with respect to the time independent background gauge field $\bar{A}$. 
The first two terms in (\ref{70}) are evidently  boundary terms. The third term is a WZW-type term written as  an integral over a $(2k+1)$ manifold, corresponding to the droplet and time. The usual 3-form in the integrand of the WZW-term, $(G^\dagger d G)^3$, has now been augmented to the appropriate $(2k+1)$-form $(G^\dagger d G)^3 \wedge \Omega^{k-1}$. Since the WZW-term is the integral of a locally exact form \cite{KN3}, the whole action $S_0$ should be considered as part of the edge action. 

\subsection{Calculation of $S_A$}

The part of the action which depends on the external gauge field $A_\mu$ is given by
\beq
S_A= - {N \over N'}   \int dtd\mu ~\tr  \left[  \rho_0 *U^\dagger * \A *U \right]
\label{72}
\eeq
where $\A$ is given in (\ref{66}). Using (\ref{23a})-(\ref{30a}) and the expression (\ref{47}) for the star-product we find that
\beq
S_A = - {N \over N'} \int dt d\mu \left[ \rho_0 ~\tr \A +{1 \over n} P^{ij} \del_i \rho_0~ \tr D_j \A -{i \over n} (\Omega^{-1})^{ij} \del_i \rho_0~ \tr (D_j G G^\dagger \A) \right]
\label{73}
\eeq
Expression (\ref{73}) naturally splits into two parts. A term that expresses the coupling between the external gauge field $A_\mu$ and the matter field $G$ and a term that involves only the gauge field $A_\mu$. In particular we find, up to $1/n^2$ order terms, 
\beq
S_A = S_{A,~{\rm matter}} + S_{A,~{\rm pure}}
\label{74}
\eeq
where
\beqar
S_{A,~{\rm matter}} & = & {N \over {nN'}} \int dt d\mu ~\del_i \rho_0 (\Omega^{-1})^{ij} \tr \left[ i(A_0 + u^k A_k) D_j G G^\dagger \right] \nonumber \\
& = & -{N \over {nN'}} \int dt d\mu ~\rho_0 (\Omega^{-1})^{ij} \tr ~\del_i \left[ i(A_0 + u^k A_k) D_j G G^\dagger \right] 
\label{75}
\eeqar
and 
\beqar
S_{A,~{\rm pure}} & = &-{N \over N'} \int dt d\mu ~\rho_0 \tr \Big[  A_0 + u^k A_k + { 1 \over {2n}} (\Omega^{-1})^{ij}  A_i\left( 2 D_jA_0 - \del_0 A_j + i [A_j,~A_0] \right) \nonumber \\
&& + {1 \over {2n}} (\Omega^{-1})^{ij} \left[ A_i A_k  \del_j u^k +  A_i \left(  2 D_j A_k - D_k A_j + i [ A_j,~A_k] + 2 \bar{F}_{jk} \right) u^k \right] \nonumber \\
&& + {1 \over {2n^2}} g^{ik} (\Omega^{-1})^{jl} \tr \left( \nabla_i A_j + \nabla_j A_i \right) \nabla_k\del_l V \Big] \nonumber \\
&& - {N \over {nN'}} \int dt d\mu ~ \del_i \rho_0 ~g^{ij}\tr \del_j ( A_0 + u^k A_k)
\label{76}
\eeqar

In doing partial integrations as in (\ref{75}), we used, along with the fact that $\rho_0$ is time independent, the relation
\beq
\del_i \left( (\Omega^{-1})^{ij} \sqrt{\det \Omega}\right) = 0
\label{68b}
\eeq

From expressions (\ref{75}) and (\ref{76}) we notice that $S_{A, {\rm matter}}$ contributes to the edge action as expected, since the matter field $G$ resides on the edge and describes the edge excitations of the droplet, while $S_{A,~{\rm pure}}$ contributes to both the edge and bulk action. 

Let us for now focus on the ``topological" part of $S_{A,{\rm pure}}$, namely the terms which contain $(\Omega^{-1})^{ij}$ but not explicitly the metric. 
\beqar
S_{A,{\rm pure}}^{\rm topological}  &= & -{N \over {N'}} \int dt d\mu \rho_0 \tr \Big[ A_0 + u^kA_k +{ 1 \over 2n} (\Omega^{-1})^{ij}  \Big[   A_i\left( 2 D_jA_0 - \del_0 A_j + i [A_j,~A_0] \right) \nonumber \\
&&+ A_i A_k  \del_j u^k +  A_i \left(  2 D_j A_k - D_k A_j + i [ A_j,~A_k] + 2 \bar{F}_{jk} \right) u^k \Big] \Big] \nonumber \\
&=& -{N \over {N'}} \int dt d\mu \rho_0  \tr \Big[ A_0 + u^kA_k +{ 1 \over 2n}(\Omega^{-1})^{ij}   \Big[   A_i  D_jA_0 +A_0 D_i A_j +A_jD_0A_i  + i A_i [A_j,~A_0]  \nonumber \\
&&+ \left( A_i  D_j A_k + A_kD_i A_j+A_jD_k A_i + i A_i [ A_j,~A_k] + 2 A_i \bar{F}_{jk} \right) u^k \nonumber \\
&&+ \del_j \left( A_i (A_0 + u^k A_k) \right)\Big] \Big]
\label{100}
\eeqar
Using (\ref{dmuo}) and 
\beq
d\mu \left[ (\Omega^{-1})^{ij} (\Omega^{-1})^{kl} + (\Omega^{-1})^{ki} (\Omega^{-1})^{jl} (\Omega^{-1})^{jk} (\Omega^{-1})^{il} \right]  =  4k(k-1) \epsilon^{ijkli_3j_3\cdots i_kj_k} \Omega_{i_3j_3} \cdots \Omega_{i_kj_k} {{d^{2k}x} \over {(4\pi)^k}} 
\label{78}
\eeq
we can rewrite the terms in (\ref{100}) as follows:
\beqar
&&-{N \over {2nN'}} \int dt d\mu \rho_0 (\Omega^{-1})^{ij}  \tr    \left( A_i  D_jA_0 +A_0 D_i A_j +A_jD_0A_i  + i A_i [A_j,~A_0] \right)  \nonumber \\
&&= {kN \over {4 \pi n N'}} \int \rho_0 \tr \left( ADA + {2i \over 3} A^3 \right) \wedge \left( {\Omega \over 2\pi} \right)^{k-1} \nonumber \\
&& = {kN \over {4 \pi n N'}} \int \rho_0 \tr \left( (A+\bar{A}) d (A+ \bar{A}) + {2i \over 3} (A+\bar{A})^3 \right) \wedge \left( {\Omega \over 2\pi} \right)^{k-1} \nonumber \\
&&-{N \over {2nN'}} \int dt d\mu \rho_0 (\Omega^{-1})^{ij} \tr ~\del_i  (A_0 \bar{A}_j)
\label{101}
\eeqar
In deriving the last expression in (\ref{101}) we used the relation
\beq
\tr \left[ (A+\bar{A}) d (A+ \bar{A}) + {2i \over 3} (A+\bar{A})^3 \right] = \tr \left[ ADA +{2i \over 3} A^3 + 2A \bar{F} -d (\bar{A} A) \right]
\label{102}
\eeq
and the fact that $(\Omega^{-1})^{ij} \bar{F}^a_{ij} =0$.
We similarly find that 
\beqar
&& -{N \over {2nN'}} \int dt d\mu \rho_0 (\Omega^{-1})^{ij}  \tr \left[ A_i  D_j A_k + A_kD_i A_j+A_jD_k A_i + i A_i [ A_j,~A_k] + 2 A_i \bar{F}_{jk} \right] u^k \nonumber \\
&=& -{N \over {nN'}} {{2k(k-1)} \over {(4 \pi)^2}} \int dt \rho_0 \left( ADA +{2i \over 3} A^3 +2 A \bar{F} \right) dV \wedge \left( {\Omega \over 2\pi} \right)^{k-2}  \label{103} \\
&=& -{N \over {nN'}} {{2k(k-1)} \over {(4 \pi)^2}} \int dt \rho_0 \left[ \left( (A+\bar{A})d(A+ \bar{A}) +{2i \over 3} (A+\bar{A})^3  \right) dV + d(\bar{A} A) dV \right] \wedge \left( {\Omega \over 2\pi} \right)^{k-2} \nonumber
\eeqar
Upon partial integration the last term in (\ref{103}) is zero for $V=V(r^2)$ and $\rho_0 = \rho_0(r^2)$.

Combining (\ref{76}), (\ref{100}), (\ref{101}) and (\ref{103}) we can rewrite $S_{A, {\rm pure}}$ in the following way
\beqar
S_{A, {\rm pure}} & = & -{N \over {N'}} \int dt d\mu \rho_0 \tr \left(A_0 + u^k A_k \right) \nonumber \\
&&+ {kN \over {4 \pi n N'}} \int \rho_0 \tr \left( (A+\bar{A}) d (A+ \bar{A}) + {2i \over 3} (A+\bar{A})^3 \right) \wedge \left( {\Omega \over 2\pi} \right)^{k-1} \nonumber \\
&&-{N \over {nN'}} {{2k(k-1)} \over {(4 \pi)^2}} \int dt \rho_0  \tr \left( (A+\bar{A})d(A+ \bar{A}) +{2i \over 3} (A+\bar{A})^3  \right) dV  \wedge \left( {\Omega \over 2\pi} \right)^{k-2} \nonumber \\
&&+ {N \over {2nN'}} \int dt d\mu \del_i \rho_0 (\Omega^{-1})^{ij}\tr  \left[ - (A_0 + u^k A_k) A_j + A_0 \bar{A}_j \right] \nonumber \\
&& - {N \over {nN'}} \int dt d\mu \del_i \rho_0 g^{ij} \tr  \del_j ( A_0 + u^k A_k) \nonumber\\
&& -{N \over {2n^2N'}} \int dt d\mu \rho_0 g^{ik} (\Omega^{-1})^{jl} \tr \left( \nabla_i A_j + \nabla_j A_i \right) \nabla_k\del_l V  
\label{104}
\eeqar
Doing a partial integration and using the fact that $\nabla_i g^{jk} = \nabla_i (\Omega^{-1})^{jk} = 0 $ and $\tr \nabla_i A_j = \tr ( \nabla_j A_i + F_{ij} ) $, we can rewrite the last term in (\ref{104}) in the following way:
\beqar
&&- {N \over {2n^2 N'}} \int dt d \mu \rho_0 g^{ik} (\Omega^{-1})^{jl} \tr \left( \nabla_i A_j + \nabla_j A_i \right) \nabla_k\del_l V \nonumber \\
& &=  {N \over {2nN'}} \int dt d \mu \del_i \rho_0 g^{ij} \tr \left( \nabla_j A_k + \nabla_k A_j \right) u^k \nonumber \\
&&+ {N \over {2nN'}} \int dt d \mu \rho_0 g^{ik} \tr \Big[    \nabla_k F_{ij} u^j - { 1 \over n}  (\Omega^{-1})^{jl}  [ \nabla_j,~\nabla_l ] A_i ~\del_k V \Big]
\label{76a}
\eeqar
The last term in (\ref{76a}) involves the curvature of ${\bf CP}^k$. Using  (\ref{46}) and (\ref{56a}) we find that 
\beqar
[\nabla_j,~\nabla_l] A_i & = & R_{jlim} A^m \nonumber \\
R_{ijlm} & = & - \left( f^{a\alpha \beta} \bar{F}^a_{ik} + {{k+1} \over k} \epsilon^{\alpha\beta} \Omega_{ik} \right) E_j^\alpha E_l^{\beta} 
\label{200}
\eeqar
Using (\ref{200}) we find that the last term in (\ref{76a}) can be written as 
\beq
 -{N \over {2n^2N'}} \int dt d \mu \rho_0 g^{ik} (\Omega^{-1})^{jl}  \tr [ \nabla_j,~\nabla_l ] A_i ~\del_k V = {{N(k+1)}\over {2nN'}} \int dt d \mu \rho_0 \tr A_k u^k
\eeq
We are now ready to write down the final edge and bulk contributions 
resulting from $S_A$ in (\ref{73}):
\beqar
S_{A}^{{\rm edge}} && =  {N \over {2nN'}} \int dt d\mu  \del_i \rho_0 (\Omega^{-1})^{ij} \tr \Big[ 2 i (A_0 + u^k A_k) D_j G G^\dagger  -
  (A_0 + u^k A_k) A_j + A_0 \bar{A}_j \Big] \nonumber \\
&& - {N \over {2nN'}} \int dt d\mu \del_i \rho_0~g^{ij}  \tr \Big[ 2 \del_j  ( A_0 + u^k A_k) - \left( \nabla_j A_k + \nabla_k A_j \right) u^k  \Big]
\label{77}
\eeqar
\beqar
S_{A}^{\rm bulk}  & = & -{N \over {N'}} \int dt d\mu \rho_0 \tr \left(A_0 + u^k A_k \right) \nonumber \\
&&+ {kN \over {4 \pi n N'}} \int \rho_0 \tr \left( (A+\bar{A}) d (A+ \bar{A}) + {2i \over 3} (A+\bar{A})^3 \right) \wedge \left( {\Omega \over 2\pi} \right)^{k-1} \nonumber \\
&&-{N \over {nN'}} {{2k(k-1)} \over {(4 \pi)^2}} \int dt \rho_0 \left[ \left( (A+\bar{A})d(A+ \bar{A}) +{2i \over 3} (A+\bar{A})^3  \right) dV \right] \wedge \left( {\Omega \over 2\pi} \right)^{k-2} \nonumber \\
&& +{N \over {2nN'}} \int dt d\mu \rho_0 \tr \Big[    \nabla^i F_{ik}   + (k+1)  A_k
\Big] u^k
\label{77a}
\eeqar
Let us momentarily focus on $S_A^{\rm bulk}$. In terms of the dimensionful coordinates $\tilde{x}$, the last term is of order $1/(BR^2)$ while the term before last is of order $1/B$ but contains higher derivatives of the external field compared to the other terms in (\ref{77a}). If we also consider the approximation where $R$ becomes large and the gradients of the external field are small compared to $B$, the last two terms become subdominant compared to the other terms in (\ref{77a}). 

A similar expression for $S_A^{\rm bulk}$, when $\rho_0=1$ (completely filled LLL level), has also been derived in \cite{nair} using a different analysis. In fact it was shown there that the ``topological" part of (\ref{77a}), where the last two terms are neglected, can be written as a single $(2k+1)$-dimensional Chern-Simons term to all orders in $1/n$. Using (78) and the fact that $N/N' = n^k /k!$ at large $n$, we can rewrite the ``topological" part of $S_A^{\rm bulk}$ (when $\rho_0=1$) as
\beqar
S_{A}^{\rm bulk}  & = &{ { (-1)^{k+1} }\over {(2\pi)^k k!}} \int  \biggl[ \tr A  \wedge \left( -n \Omega \right)^k \nonumber \\
&&+ {k \over 2}~ \tr \left( (A+\bar{A}+V) d (A+ \bar{A}+V) + {2i \over 3} (A+\bar{A}+V)^3 \right) \wedge \left( -n \Omega \right)^{k-1} \nonumber \\
&&+{{k(k-1)} \over 2}~ \tr \left( (A+\bar{A})d(A+ \bar{A}) +{2i \over 3} (A+\bar{A})^3  \right) dV  \wedge \left( -n \Omega  \right)^{k-2} \biggr]
\label{77b}
\eeqar
One can check, using that $da=n \Omega $, that up to a constant term independent of the fluctuating field, this is indeed the large $n$ expansion of the $(2k+1)$-dimensional Chern-Simons term for the gauge field $\tilde{A}$:
\beqar
S_A^{\rm bulk} & = & S_{CS} (\tilde{A}) \nonumber \\
\tilde{A} & = & \Big( A_0 +V,~ -a_i+\bar{A}_i+A_i \Big)
\eeqar
in agreement with \cite{nair}.\footnote{In comparing with \cite{nair} one has to take $a \rightarrow -a$ or equivalently $n \Omega \rightarrow -\omega$. The extra $(-1)^{k+1}$ factor in front  of the Chern-Simons action in (\ref{77b}) has to do with the fact that, in our notation, the components of the gauge fields are related to the matrix form by $A= A^a (T_a)^T$ whereas \cite{nair}
used the definition $A= A^a (-T_a)^T$}

\section{ Anomaly cancellation between bulk and edge actions}

As we have explained in section 2, the full bosonic action $S$ is by construction invariant under
\beq
\delta U  =  -i \lambda * U 
\label{79a}
\eeq
and
\beq
\delta A_\mu  =  D_\mu \Lambda + i [A_\mu,~ \Lambda] 
\label{79}
\eeq
which further induces the $W_N$ transformation (\ref{11}).

Using (\ref{23a})-(\ref{30a}) one can show that (\ref{79a}) implies the following gauge transformation for $G$:
\beq
\delta G~G^\dagger = -i \Lambda + \cdots 
\label{80}
\eeq
where $\cdots$ indicates higher order terms in $1/n$.
This means that the total effective action we derived to $1/n$ order,
\beq
S = (S_0+ S_A)^{\rm edge} + S_A^{\rm bulk}
\label{81}
\eeq
is automatically gauge invariant under (\ref{79}) and (\ref{80}).  One can verify this explicitly by calculating the gauge variation of $S$. In fact it is interesting to consider separately the gauge transformation of the edge and bulk parts of the action. 

\subsection{Gauge transformation of $S^{\rm edge}$}

The total edge action is 
\beqar
S^{\rm edge} & = & S_0 + S_A^{\rm edge} \nonumber \\
& = & {N \over {2nN'}} \int  dt d\mu ~\del_i  \rho_0 (\Omega^{-1})^{ij} 
~ \tr \Big[ \left( G^{\dagger} {\dot G} + u^k G^{\dagger} D_k G \right)  
G^{\dagger}D_j G   - i \left( {\dot G} G^\dagger + G^\dagger{ \dot G} \right) \bar{A}_j \nonumber\\
&& + 2i (A_0 + u^k A_k) D_j G G^{\dagger} - ( A_0 + u^k A_k) A_j + A_0 \bar{A}_j \Big] \nonumber \\
&&-{{N} \over {2nN'}} \int dt d\mu \rho_0   (\Omega^{-1})^{ij} \tr \left( G^{\dagger}{\dot G} 
 G^{\dagger }\del_iG ~G^{\dagger }\del_jG \right) \nonumber \\
 && -{{N} \over {2nN'}} \int dt d\mu \del_i \rho_0   g^{ij} \tr \left[ 2 \del_j (A_0 + u^k A_k) - (\nabla_j A_k + \nabla_k A_j) u^k \right] 
\label{150}
\eeqar
The last two terms do not involve matter coupling; they further depend explicitly on the metric and are of no topological nature. One can show that they are gauge invariant by making use of the K\"ahler property of the manifold, namely 
\beq
g^{z\bar{z}} = g^{\bar{z}z} = \del_z \del_{\bar{z}} K (r^2)
\label{83}
\eeq
where $K(r^2)$ is the K\"ahler potential. The rest of $S^{\rm edge}$ transforms in the following way under the gauge transformation (\ref{79}) and (\ref{80}) :
\beq
\delta S^{\rm edge}  =  {N \over {2nN'}} \int d\mu \del_i \rho_0 (\Omega^{-1})^{ij} \tr \left[ A_0 \del_j \Lambda - A_j \del_0 \Lambda \right] 
\label{84}
\eeq
The gauge variation of $S^{\rm edge}$ does not depend on $u^k$. In fact it is interesting to write down the $u$-independent ($V=0$) part of the edge action neglecting the last metric dependent term. After some rearrangement of the terms we find that this can be written as a higher dimensional WZW action, gauged in a left-right asymmetric way as follows:
\beqar
S^{\rm edge}(u^k=0) &=&S_{WZW} ( A^L = A + \bar{A}, A^R= \bar{A}) \label{81} \\
&=& {N \over {2nN'}} \int dt d\mu \del_i \rho_0 (\Omega^{-1})^{ij} G^{\dagger} (\del_0 + i A_0^LG -iG A_0^R)~G^{\dagger} (\del_j G +i A_j^L G -iG A_j^R) \nonumber \\
&+&{Nk \over {4 \pi nN'}} \int \rho_0 \left[ -d \left( iA^L dG G^{\dagger} +i A^R G^{\dagger} dG + A^L G A^R G^{\dagger} \right) +{1 \over 3} \left( G^{\dagger} dG \right)^3 \right] \wedge \left( {\Omega \over 2\pi} \right) ^{k-1} \nonumber 
\eeqar
The first term in (\ref{81}) is gauge invariant, while the last two terms combine to produce
\beq
\delta S^{\rm edge}(u^k=0) = {Nk\over {4\pi nN'}} \int  d \rho_0 ~ \tr ( d A^L  \Lambda) \wedge \left( {\Omega \over 2\pi} \right)^{k-1} \label{81a}
\eeq
which is the same as (\ref{84}) for a time independent $\bar{A}_i$, with $\bar{A}_0=0$.
The full $S^{\rm edge}$ action, including the $u^k$ dependent terms can also be written as a gauged WZW action by doing the following substitutions in (\ref{81}):
\beqar
& \del_0 \rightarrow \del_{\tau}= \del_0 + u^k \del_k    & \nonumber \\
& A_0^L \rightarrow A^L_{\tau} = A_0 + u^k (A_k + \bar{A}_k) &A_0^R \rightarrow A^R_{\tau} = u^k \bar{A}_k \nonumber \\
& A_i^L = A_i + \bar{A}_i & A^R_i = \bar{A}_i 
\label{81b}
\eeqar
One can explicitly verify that the $u$-dependent terms are gauge invariant. The derivative $\del_{\tau}$ is the convective derivative, well known in hydrodynamics. The appearance of $A_\tau$ is consistent with the gauging of the convective derivative.

\subsection{Gauge transformation of $S^{\rm bulk}$}

The full bulk action $S^{\rm bulk}$ is given by $S_A^{\rm bulk}$ in (\ref{77a}). The gauge transformation of the combination $\tr(A_0+u^k A_k)$ is
\beq
\delta \tr (A_0+u^k A_k) = \tr \left( \dot{\Lambda} + { 1\over n} { {\del V} \over {\del r^2}} \del_{\Omega} \Lambda \right) 
\eeq
where $\del_{\Omega} = (\Omega^{-1})^{ij} 2 \hat{x}_j \del_i$ is an angular derivative along the boundary. 
Using the fact the $\rho_0$ is time independent and that $\rho_0,~V$ are spherically symmetric, one can easily show by partial integration that the terms in the first and last line of (\ref{77a}) are gauge invariant. 
Further
\beq
\delta \tr \left[ (A + \bar{A}) d (A + \bar{A}) + { 2i \over 3} (A + \bar{A})^3 \right] = d ~\tr \left( \Lambda d (A + \bar{A}) \right)
\label{90}
\eeq
which implies that the $V$-dependent Chern-Simons term in the third line of (\ref{77a}) is gauge invariant. The gauge non-invariance of the bulk action $S^{\rm bulk}$ is due to the $V$-independent K\"ahler-Chern-Simons term in the second line. In fact we find
\beq
\delta S^{\rm bulk}  =  - {Nk \over {4 \pi nN'}} \int  d \rho_0  ~ \tr ~ \left[ d( A+\bar{A} ) \Lambda \right]  \wedge \left( {\Omega \over 2\pi} \right)^{k-1} \label{86}
\eeq
Adding the gauge variations of the edge and bulk actions we find, as expected, that the total bosonic action $S$ is gauge invariant,
\beq
\delta S = \delta S^{\rm edge}  + \delta S^{\rm bulk} = 0
\label{87}
\eeq

\section{ Summary, concluding remarks}

In this paper we derived an exact bosonic action describing the dynamics of the LLL fermions on ${\bf CP}^k$ in the presence of non-Abelian gauge interactions. It is a one-dimensional gauged matrix action written in terms of $(N \times N)$ matrices acting on the $N$-dimensional lowest Landau level single particle Hilbert space and can be further expressed as an action of a noncommutative field theory. Its semiclassical limit, as $N \rightarrow \infty$ and the number of fermions becomes large, produces an effective action describing the gauge interactions of a $\nu = 1$ higher dimensional, non-Abelian  quantum Hall droplet. The effective action contains a bulk contribution in the form of Chern-Simons type actions in terms of the external gauge fields and a boundary contribution in terms of a higher dimensional gauged, chiral Wess-Zumino-Witten action. The gauging of the boundary Wess-Zumino-Witten action appears in a left-right asymmetric way, such that there is an exact anomaly cancellation between the bulk and the boundary terms, guaranteeing the gauge invariance of the total action. Further the bulk Chern-Simons type terms can be combined into a single $(2k+1)$-dimensional Chern-Simons action for the total gauge field, including the Abelian and non-Abelian background and fluctuating gauge fields as well as the confining potential, in agreement with \cite{nair}. 

Given the fact that the full single particle Hilbert space for all Landau levels is known in the case of ${\bf CP}^k$ \cite{KN2,KN3}, our results can be extended to derive the effective action for a $\nu =n$ higher dimensional Abelian or non-Abelian  quantum Hall droplet, where $n$ Landau levels are filled.  Further extensions towards including inter-Landau level transitions and interactions is interesting and worth pursuing. 

Although many of the explicit calculations in this paper were done in the context of the QHE formulation on ${\bf CP}^k$, the outlined bosonization procedure is quite general and applies to any manifold which admits a consistent formulation of QHE. Furthermore, since 
the lowest Landau level of a $2k$-dimensional nonrelativistic fermionic system can also be thought of as the phase space of a lower $k$-dimensional system, our analysis can be clearly interpreted as a phase space bosonization of $k$-dimensional fermions. Related work on phase space Hall droplets has been done in \cite{poly1, boso6}.

Implications of this work to higher dimensional fluids is also of interest.
\vskip .2in
\noindent
{\bf Acknowledgements}

I would like to thank V.P. Nair for many useful discussions and A.P. Polychronakos for comments on the manuscript. This work was supported in part by the National Science Foundation under grant number PHY-0457304 and a PSC-CUNY grant.


\begin{thebibliography}{99}

\bibitem{HZ} S.C. Zhang, J.P. Hu, {\it Science} {\bf 294} (2001) 823; 
J.P. Hu, S.C. Zhang, cond-mat/0112432.

\bibitem{KN1} D. Karabali, V.P. Nair, {\it Nucl. Phys.} {\bf B641} (2002) 533.

\bibitem{KN2} D. Karabali and V.P. Nair, {\it Nucl. Phys.} {\bf B679} (2004) 427.

\bibitem{KN3} D. Karabali and V.P. Nair, {\it Nucl. Phys.} {\bf B697} (2004) 513.

\bibitem{KN4} D. Karabali, V.P. Nair and S. Randjbar-Daemi, hep-th/0407007, {\it Fuzzy spaces, the M(atrix) model and the quantum Hall effect}, in Ian Kogan memorial volume, ``From Fields to Strings: Circumnavigating Theoretical Physics", ed. M. Shifman, A. Vainshtein and J. Wheater.

\bibitem{KA} D. Karabali, {\it Nucl. Phys.} {\bf B726} (2005) 407.

\bibitem{everyone} M. Fabinger, {\it JHEP} {\bf 0205} (2002) 037; 
Y.X. Chen, B.Y. Hou, B.Y. Hou, {\it Nucl. Phys.} {\bf B638} (2002)
220; Y. Kimura, {\it Nucl. Phys.} {\bf B637} (2002) 177; H. Elvang, J. Polchinski, hep-th/0209104; B.A. Bernevig, C.H. Chern, J.P. Hu, N. Toumbas, S.C. Zhang, 
{\it Ann. Phys.} {\bf 300} (2002) 185;
B. A. Bernevig, J.P. Hu, N. Toumbas, S.C. Zhang, {\it Phys. Rev. Lett.}
{\bf 91} (2003) 236803; 
S.C. Zhang, {\it Phys. Rev. Lett.} {\bf 90} (2003) 196801; B. Dolan, {\it JHEP} {\bf 0305} (2003) 18; G. Meng, {\it J. Phys.} {\bf A36} (2003) 9415; S. Bellucci, P.Y. Casteill and A. Nersessian, {\it Phys. Lett.} {\bf B574} (2003) 121; V.P. Nair and S. Randjbar-Daemi, {\it Nucl. Phys.}
{\bf B679} (2004) 447; A. Jellal, {\it Nucl. Phys.} {\bf B725} (2005) 554.
  
\bibitem{poly1} A.P. Polychronakos, {\it Nucl. Phys.} {\bf B705} (2005) 457; {\it Nucl. Phys.} {\bf B711} (2005) 505.


\bibitem{wen} X.G. Wen, {\it Phys. Rev. } {\bf B41} (1990) 12838; D.H. Lee and X.G. Wen, {\it Phys. Rev. Lett.} {\bf 66} (1991) 1765; M. Stone, {\it Phys. Rev.} {\bf B42} (1990) 8399; {\it Ann. Phys.} (NY) {\bf 207} (1991) 38; J. Frohlich and T. Kepler, {\it Nucl. Phys.} {\bf B354} (1991) 369.


\bibitem{IKS} S. Iso, D. Karabali and B. Sakita, {\it Phys. Lett.} { \bf B296} (1992) 143; A. Cappelli, G. Dunne, C. Trugenberger and G. Zemba, {\it Nucl. Phys.} {\bf B398} (1993) 531; A. Cappelli, C. Trugenberger and G. Zemba, {\it Nucl. Phys.} {\bf B396} (1993) 465; {\it Phys. Rev. Lett.} {\bf 72} (1994) 1902; D. Karabali, {\it Nucl. Phys.} {\bf B419} (1994) 437; {\it Nucl. Phys.} {\bf B428} (1994) 531; M. Flohr and R. Varnhagen, {\it J. Phys.} {\bf A27} (1994) 3999.

\bibitem{boso1} A. Luther, {\it Phys. Rev.} {\bf B19} (1979) 320.

\bibitem{boso2} F.D. Haldane, {\it Helv. Phys. Acta} {\bf 65} (1992) 152; cond-mat/0505529.

\bibitem{boso3} A.H. Castro Neto and E. Fradkin, {\it Phys. Rev.} {\bf B49} (1994) 10877.

\bibitem{boso4} A. Houghton and B. Marston, {\it Phys. Rev.} {\bf B48} (1993) 7790; H.J. Kwon, A. Houghton and B. Marston, {\it Phys. Rev.} {\bf B52} (1995) 8002.

\bibitem{boso5} P.W. Anderson and D. Khveshchenko, {\it Phys. Rev.}{\bf B52} (1995) 16415.

\bibitem{boso6} A. P. Polychronakos, hep-th/0502150.

\bibitem{boso7} A. Dhar, G. Mandal and N.V. Suryanarayana, {\it JHEP} {\bf 0601} (2006) 118; A. Dhar and G. Mandal, hep-th/0603154.

\bibitem{wadia} S.R. Das, A. Dhar, G. Mandal and S.R. Wadia, {\it Int. J. Mod. Phys.} {\bf A7} (1992) 5165;    
{\it Mod. Phys. Lett.} {\bf A7} (1992) 71; A. Dhar, G. Mandal and S.R. Wadia, {\it Int. J. Mod. Phys.} {\bf A8} (1993) 325; {\it Mod. Phys. Lett.} {\bf A7} (1992) 3129; {\it Mod. Phys. Lett.} {\bf A8} (1993) 3557; A. Dhar, {\it JHEP} {\bf 0507} (2005) 064.


\bibitem{sakita1} B. Sakita, {\it Phys. Lett.} {\bf B387} (1996) 118;
B. Sakita and R. Ray, {\it Phys. Rev.} {\bf B65} (2001) 035320.  


\bibitem{sakita2} B. Sakita, {\it Phys. Lett.} {\bf B315} (1993) 124.

\bibitem{shizuya} A similar approach in the case of the two-dimensional QHE, based however on $W_{\infty}$ transformations mixing higher Landau levels has been proposed by  K. Shizuya, {\it Phys. Rev.} {\bf B52} (1995) 2747.


\bibitem{seib1} N. Seiberg and E. Witten, {\it JHEP} {\bf 9909} (1999) 032.

\bibitem{seib2} J. Madore, S. Schraml, P. Schupp and J. Wess {\it Eur. Phys. J.}{\bf C16} (2000) 161; B. Jurco, L. Moller, S. Schraml, P. Schupp and J. Wess{\it Eur. Phys. J.} {\bf C21} (2001) 383; W. Behr and A. Sykora, {\it Nucl.Phys.} {\bf B698} (2004) 473.

\bibitem{haldane} F.D.M. Haldane, {\it Phys. Rev. Lett.} {\bf 51} (1983) 605.

\bibitem{bal} G. Alexanian, A.P. Balachandran, G. Immirzi, B. Ydri, {\it J. Geom. Phys.} {\bf42} (2002) 28; A.P. Balachandran, B.P. Dolan, J. Lee, X. Martin, D. O'Connor, {\it J. Geom. Phys.}{\bf 43} (2002) 184.

\bibitem{nepomechie} L.S. Brown, R.I. Nepomechie, {\it Phys. Rev.} {\bf D35} (1987) 3239; D. Karabali, Q.H. Park, H.J. Schnitzer, Z. Yang, {\it Phys. Lett.}{\bf B216} (1989) 307; D. Karabali, H.J. Schnitzer, {\it Nucl.Phys.} {\bf B329} (1990) 649.

\bibitem{nair} V.P. Nair, hep-th/0605007.


\end{thebibliography}
\end{document}